\documentclass[12pt]{article}
\usepackage{color}
\usepackage{latexsym}
\usepackage{epsfig,amssymb,euscript, mathrsfs}
\usepackage{amsmath}
\usepackage{color}
\definecolor{MyDarkBlue}{rgb}{0.15,0.15,0.45}
\usepackage[linktocpage=true]{hyperref}
\hypersetup{
colorlinks=true,
citecolor=MyDarkBlue,
linkcolor=MyDarkBlue,
urlcolor=MyDarkBlue,
pdfauthor={Luis F Alday, Dario Martelli, Paul Richmond, James Sparks},
pdftitle={Localization and Three-manifolds},
pdfsubject={hep-th}
}

\usepackage[numbers,sort&compress]{natbib}
\newsavebox{\ns}
\newsavebox{\dbrane}
\newsavebox{\dbshort}

\def\be{\begin{equation}}
\def\ee{\end{equation}}
\def\bea{\begin{eqnarray}}
\def\eea{\end{eqnarray}}

\newcommand{\nn}{\nonumber}

\newcommand\R{\mathbb{R}}

\newcommand\C{\mathbb{C}}

\newcommand\diff{\mathrm{d}}

\newcommand{\dd}{\mathrm{d}}

\newcommand{\ii}{\mathrm{i}}

\newcommand{\ex}{\mathrm{e}}

\newcommand{\rc}{r}

\newlength{\sswidth}
\newcommand{\sla}[1]{
   \settowidth{\sswidth}{$#1$}
   \mbox{$\not{\hspace*{-0.3\sswidth}#1}$}}

\DeclareMathOperator{\image}{\mathrm{Im}\,}

\numberwithin{equation}{section}       

\begin{document}

\begin{titlepage}

\begin{center}

\today

\vskip 2.3 cm 

\vskip 5mm

{\Large \bf Localization on Three-Manifolds}

\vskip 15mm

{Luis F. Alday${}^{\,a}$, Dario Martelli${}^{\,b}$, Paul Richmond${}^{\,a}$ and James Sparks${}^{\,a}$}

\vspace{1cm}
\centerline{${}^a${\it Mathematical Institute, University of Oxford,}}
\centerline{{\it 24-29 St Giles', Oxford, OX1 3LB, UK}}
\vspace{.8cm}
\centerline{${}^b${\it Department of Mathematics, King's College London,}}
\centerline{{\it The Strand, London WC2R 2LS, UK}}

\end{center}

\vskip 2 cm

\begin{abstract}
\noindent  
We consider supersymmetric gauge theories on Riemannian three-manifolds with the topology of a three-sphere. The three-manifold is always equipped with an almost contact structure and an associated Reeb vector field. We show that the partition function depends only on this vector field, giving an explicit expression in terms of the double sine function. In the large $N$ limit our formula agrees with a recently discovered two-parameter family of dual supergravity  solutions. 
We also explain how our results may be applied to prove vortex-antivortex factorization. Finally, we comment on the extension of our results to three-manifolds with non-trivial fundamental group.
\end{abstract}

\end{titlepage}

\pagestyle{plain}
\setcounter{page}{1}
\newcounter{bean}
\baselineskip18pt
\tableofcontents


\section{Introduction}

Over the last few years there has been remarkable progress in the computation of exact quantities in supersymmetric gauge theories using localization techniques \cite{Pestun:2007rz,Kapustin:2009kz}. These exact results are important as they allow for non-perturbative tests of various conjectured dualities. 

Three-dimensional supersymmetric gauge theories on curved backgrounds have received a lot of attention recently. Explicit results are known for the round sphere \cite{Kapustin:2009kz,Jafferis:2010un,Hama:2010av},  particular squashed spheres \cite{Hama:2011ea,Imamura:2011wg} and Lens spaces \cite{Benini:2011nc,Gang:2009wy}. A uniform treatment of rigid supersymmetric theories in curved backgrounds was started in \cite{Festuccia:2011ws} and developed further for three-dimensional theories on Riemannian manifolds in \cite{Closset:2012vp,Closset:2012vg,Closset:2012ru,Klare:2012gn}. In particular, \cite{Closset:2012ru} constructed ${\cal N}=2$ supersymmetric gauge theories with an R-symmetry on Riemannian three-manifolds, including all previously studied examples. In the ``real'' case, which will be the focus of this paper,  there exist two supercharges $\epsilon$,  $\epsilon^c$, 
with opposite R-charge, and the three-manifold is equipped with an almost contact metric structure with an associated Killing Reeb vector field $K$.

The aim of this paper is to compute the partition function for ${\cal N}=2$ Chern-Simons theories, coupled to arbitrary matter, on these general backgrounds. In the case of trivial fundamental group, so $M_3$ has the topology of $S^3$, there are no non-trivial flat connections and the localized partition function reduces to a finite-dimensional integral over the Cartan of the gauge group. Either all the orbits of $K$ close, in which case $M_3$ is equipped with a 
$U(1)$ action, or there is at least a $U(1)\times U(1)$ symmetry. In the latter case $M_3\cong S^3$ 
has a \emph{toric} almost contact structure, meaning that we may write the Killing Reeb vector as
\bea
\label{Ku1u1}
K &=& b_1\partial_{\varphi_1}+b_2\partial_{\varphi_2}~,
\eea
where $\varphi_1$, $\varphi_2$ are $2\pi$-period coordinates on the torus $U(1)\times U(1)$, which 
acts in the standard way on $S^3$. In this case the partition function for a Chern-Simons-matter theory is given by
\bea
\label{Z}
Z \, = \, \int \diff\sigma_0 \, \ex^{\frac{\ii\pi k}{b_1b_2}\mathrm{Tr}\, \sigma_0^2}\prod_{\alpha\in \Delta_+} 4
\sinh \frac{\pi\sigma_0\alpha}{b_1}\sinh\frac{\pi\sigma_0\alpha}{b_2}\prod_{\rho}s_\beta\left[\frac{\ii Q}{2}(1-\rc)-\frac{\rho(\sigma_0)}{\sqrt{b_1b_2}}\right]~.
\eea
Here the integral is over the Cartan of the gauge group, $k$ denotes the Chern-Simons level, the first product is over positive roots $\alpha\in\Delta_+$ of the gauge group, and the second product is over weights $\rho$ in the weight space decomposition for a chiral matter field in an arbitrary representation $\mathcal{R}$ of the gauge group. We have also defined 
\bea
\beta &\equiv & \sqrt{\frac{b_1}{b_2}}~, \qquad Q \ \equiv \ \beta + \frac{1}{\beta}~,
\eea
the R-charge of the matter field is denoted $\rc$, and $s_\beta(z)$ denotes the double sine function. Notice that we may absorb a factor of $1/\sqrt{b_1b_2}$ into $\sigma_0$, which is then integrated over, and thus we see that the partition function (\ref{Z}) depends on the background geometry only through the single parameter $\beta^2=b_1/b_2$. We shall also present a sketch of a proof for why (\ref{Z}) continues to hold 
in the case that $K$ generates only a $U(1)$ action on $S^3$, and comment 
on the extension of our results to three-manifolds with non-trivial fundamental group.

The large $N$ limit of this class of partition functions has been considered in  \cite{Martelli:2011fu}. 
Using  \cite{Martelli:2011fu} we may compare the large $N$ limit of our field theory formula with candidate
gravity dual solutions.
In \cite{Martelli:2013aqa} the authors considered a two-parameter family of squashed sphere backgrounds, interpolating between all previously studied examples, and constructed their gravity duals. The large $N$ limit of our 
partition function for this particular case exactly agrees with the free energy of the holographic duals in  \cite{Martelli:2013aqa}.

One of the interesting properties of the formula (\ref{Z}) is that it exhibits vortex-antivortex factorization \cite{Pasquetti:2011fj}. This 
has recently been studied in detail in \cite{Beem:2012mb}, although a key step that has been missing is a direct proof of why 
the partition function on a squashed sphere should decompose into these holomorphic blocks. Our results  give a simple  
proof. The key point is that \emph{any} supersymmetric three-sphere background has partition function given by (\ref{Z}). In particular, 
we may construct such a three-sphere background by gluing together two copies of $\R^2\times_\beta S^1$, where the metric is a twisted product of a cigar metric on $\R^2$ with a circle $S^1$. 
The boundaries of these two copies of $\R^2\times_\beta S^1$ are two-tori, and we glue with an $S$-transformation to 
obtain topologically a three-sphere. Vortex-antivortex factorization is then explained by taking the limit that the cigars 
become infinitely long.\footnote{Strictly speaking the real backgrounds that we study only prove a real version 
of this factorization. We shall comment more on this later.} The point here is that the partition function depends only 
on the Reeb vector field $K$ in (\ref{Ku1u1}), which we hold fixed, and is independent of all other geometric parameters.

The rest of this paper is organized as follows. In section \ref{SecBackgroundGeometry} we discuss the background geometry of the three-manifold. In section \ref{SecSusyAndLocalization} we present the supersymmetry 
variations and derive the locus in field space onto which the partition function localizes. In sections \ref{SecClassicalAction} and \ref{SecOneLoop} we compute the classical action evaluated on this localizing locus, and the one-loop determinants, respectively. Section \ref{generalSec} contains 
comments on extending our results to more general three-manifolds. In section \ref{Applications} we discuss two applications
of our results: in 
 section \ref{LargeNLimit} we compare the large $N$ limit of our 
field theory result to the gravity duals of \cite{Martelli:2013aqa}, while in section \ref{vortex} 
we comment on how our results may be used to prove vortex-antivortex factorization.
We end with our conclusions in section \ref{SecConclusions}. Also included are two appendices
 which contain our conventions and further results regarding the background geometry.


\section{Background geometry}\label{SecBackgroundGeometry}

We study a general class of ``real'' supersymmetric backgrounds with two supercharges related to one another by charge conjugation \cite{Closset:2012vp}. If $\epsilon$ denotes the Killing spinor then there is an associated Killing vector 
\bea
K & \equiv & \epsilon^\dagger \gamma^\mu \epsilon \partial_\mu \ = \ \partial_\psi~.
\eea
This Killing vector is nowhere zero and therefore defines a foliation of the three-manifold. This foliation is transversely holomorphic with local complex coordinate $z$. In terms of these coordinates the background metric is
\bea\label{metric}
\dd s^2 &=& \Omega(z,\bar{z})^2 ( \dd \psi + a )^2 + c(z,\bar{z})^2 \dd z \dd \bar z \, ,
\eea
where $a= a (z,\bar{z}) \dd z + \overline{a (z,\bar{z} )} \dd \bar z$ is a local one-form. 

There are also two background vector fields $A_\mu$ and $V_\mu$ and a scalar function $h$. In terms of these variables the Killing spinor equation is
\bea\label{KSE}
(\nabla_\mu - \ii A_\mu) \epsilon & = & -\frac{1}{2} \ii h \gamma_\mu \epsilon -\ii V_\mu \epsilon -\frac{1}{2} \epsilon_{\mu \nu \rho} V^\nu \gamma^\rho \epsilon~.
\eea
If all these background fields, including the metric, are \emph{real} then the charge conjugate spinor satisfies (see appendix \ref{AppConventions} for our conventions)
\bea\label{KSEconj}
(\nabla_\mu + \ii A_\mu) \epsilon^c & = & -\frac{1}{2} \ii h \gamma_\mu \epsilon^c + \ii V_\mu \epsilon^c + \frac{1}{2} \epsilon_{\mu \nu \rho} V^\nu \gamma^\rho \epsilon^c~.
\eea
Here we have defined charge conjugation as
\bea
\epsilon^c & \equiv & \sigma_2 \epsilon^* \, ,
\eea
with $\sigma_2$ the second Pauli matrix.

It is  convenient to define the orthonormal frame of one-forms $e^1, e^2, e^3$ via\footnote{In \cite{Closset:2012ru} the almost contact form $\eta$ is defined as $e^3$, but we will reserve this notation for a rescaled form.}
\bea\label{frame}
e^3  & \equiv & \ \frac{\epsilon^\dagger \gamma_{(1)} \epsilon}{\epsilon^\dagger \epsilon}~, \nn \\
s (e^1+ \ii e^2) &  \equiv &  \ii \epsilon^{c \dagger} \gamma_{(1)} \epsilon \ \equiv \ P~,
\eea
where $\gamma_{(n)}\equiv \frac{1}{n!}\gamma_{a_1\cdots a_n}e^{a_1}\wedge\cdots \wedge e^{a_n}$.
The function $s=s(\psi,z,\bar{z})$ appears in the Killing spinor solution 
\bea
\epsilon &= &  \sqrt{s(\psi,z,\bar{z})} \left( \begin{array}{c} 1 \\ 0 \end{array} \right)~.
\eea
We also have that
\bea
\epsilon^\dagger \epsilon & = &  |s| \ =  \ \Omega \ = \ \Omega(z,\bar{z})~,
\eea
is nowhere zero. These quantities are related to those appearing in the metric (\ref{metric}) through
\bea
e^3 & = & \Omega(\dd\psi+a)~,\nn\\
e^1+\ii e^2 &= & c(z,\bar z) \dd z~. 
\eea
Notice that because $P = s c\, \dd z$ the integrability condition $P \wedge \dd P=0$ 
\cite{Closset:2012ru, Klare:2012gn} is satisfied, hence the functions 
$\Omega, c$ and one-form $a$ are arbitrary, subject to appropriate boundary conditions imposed
by demanding a regular three-manifold. Below we show that the other background fields are
determined in terms of the metric data, up to the following shift
transformations
\bea\label{shifts}
V_\mu &\rightarrow & V_\mu + \kappa e^3_\mu ~,\nn\\
h \ &\rightarrow & h + \kappa ~, \nn\\
A_\mu \ &\rightarrow & A_\mu + \frac{3}{2}\kappa e^3_\mu ~,
\eea
where $\kappa$ is real. Notice that a gauge-invariant combination is $A_\mu + \alpha V_\mu + \beta h e^3_\mu$ provided $\frac{3}{2} + \alpha + \beta = 0$. 
We fix the gauge by choosing the background fields $V$ and $h$ to be
\bea
V_\mu & =&  \epsilon_{\mu}^{\ \, \nu \rho} \partial_\nu e^3_\rho~,\nn\\
h & = & \frac{1}{2} e^3_\mu V^\mu \  = \  \frac{1}{2} \epsilon^{\mu \nu \rho} e^3_\mu \partial_\nu e^3_\rho~.
\eea
Then also
\bea\label{Vh}
V &=& * \, \diff e^3~, \qquad h \ = \  \frac{1}{2}* (e^3 \wedge \dd e^3)~.
\eea
Using these formulae one can easily show that
\bea\label{V1V2}
V_1 &=& \partial_2 \log \Omega~, \qquad V_2 \ = \ -\partial_1\log \Omega~,
\eea
while
\bea\label{V3}
V_3 &=& e^3_\mu V^\mu \ = \ 2h~.
\eea
We also record the following formula for $A$:
\bea\label{Aformula}
A_\mu -\frac{1}{2} h e^3_\mu - V_\mu&=&   j_\mu~,
\eea
where
\bea
j_\mu &\equiv & \frac{\ii}{4 \Omega^2} \left(s \partial_\mu \bar s - \bar s \partial_\mu s \right) + \frac{1}{2} \omega_\mu{}^1{}_2 ~.
\eea
Notice that the left hand side of (\ref{Aformula}) is invariant under the gauge symmetry (\ref{shifts}).
A derivation of this formula, together with further details on the background geometry, may be found in appendix \ref{AppMoreFormulae}.

It is worth pointing out that the discussion above implies that if a metric $g$
solves the Killing spinor equation, then any conformally related metric $\hat g
= \lambda^2 g$, with $\lambda$ nowhere vanishing, yields another solution. This is
related to the fact that equation (\ref{KSE}) is equivalent to the charged conformal
Killing spinor equation \cite{Klare:2012gn}, that is known to be conformally invariant. 


\section{Supersymmetry and localization}\label{SecSusyAndLocalization}

In this section we write the supersymmetry transformations for $\mathcal{N}=2$ supersymmetric multiplets on 
the above three-manifold backgrounds. Via a standard argument the partition function localizes onto supersymmetric configurations.

\subsection{Vector multiplet}

The $\mathcal{N}=2$ vector multiplet contains a scalar $\sigma$, gauge field $\mathcal{A}$ with 
field strength $\mathcal{F}$, gaugino $\lambda$ and D-term $D$, all in the adjoint representation of the gauge group $G$. 
The supersymmetry transformations are
\begin{align}\label{SUSYvector}
\mathcal{Q}\mathcal{A}_\mu \ =& \ -\frac{\ii}{2}\lambda^\dagger\gamma_\mu\epsilon~, & \mathcal{Q}\sigma \ \ =& \ \frac{1}{2}\lambda^\dagger\epsilon~,\nn\\
\mathcal{Q}\lambda \ =& \ \left[-(D-\sigma h)+\frac{\ii}{2}\epsilon^{\mu\nu\rho}\gamma_\rho\mathcal{F}_{\mu\nu} + (\ii{D}_\mu\sigma - V_\mu \sigma)\gamma^\mu\right]\epsilon~, & \mathcal{Q}\lambda^\dagger \ =& \ 0~,\nn\\
\mathcal{Q}D \ =& \ \frac{\ii}{2}D_\mu(\lambda^\dagger\gamma^\mu\epsilon) + \frac{1}{2}V_\mu\lambda^\dagger\gamma^\mu\epsilon 
- \frac{1}{2}h\lambda^\dagger\epsilon - \frac{\ii}{2}[\lambda^\dagger\epsilon,\sigma]~. & \ & \ 
\end{align}
Here the covariant derivatives acting on the fields are 
\bea
D_\mu \sigma &=& \nabla_\mu \sigma - \ii [\mathcal{A}_\mu,\sigma]~,\nn \\
D_\mu \lambda^\dagger &=& \nabla_\mu \lambda^\dagger - \ii [\mathcal{A}_\mu,\lambda^\dagger] + \ii \left( A_\mu - \tfrac{1}{2} V_\mu \right) \lambda^\dagger~,\nn  \\
D_\mu \epsilon &=& \nabla_\mu \epsilon - \ii \left( A_\mu - \tfrac{1}{2} V_\mu \right) \epsilon~,
\eea
and $\mathcal{F}_{\mu\nu}$ is the curvature
\bea
\mathcal{F}_{\mu\nu} & = & \partial_\mu \mathcal{A}_\nu - \partial_\nu \mathcal{A}_\mu - \ii [ \mathcal{A}_\mu , \mathcal{A}_\nu ] \, .
\eea
We will be interested in Chern-Simons-matter theories, for which the Chern-Simons action is
\bea\label{CSaction}
\mathcal{L}_{CS} &=& \frac{\ii k}{4\pi} {\rm Tr} \left[\epsilon^{\mu\nu\rho} \left( \mathcal{A}_\mu \nabla_\nu \mathcal{A}_\rho -\frac{2\ii}{3} \mathcal{A}_\mu \mathcal{A}_\nu \mathcal{A}_\rho \right) - \lambda^\dagger \lambda + 2 D \sigma \right] \, ,
\eea
and $k$ is the Chern-Simons level, which is an integer for $U(N)$ gauge groups.

The path integral localizes on bosonic backgrounds with $\mathcal{Q}\lambda=0$. This  reads
\bea
0 &=& \mathcal{Q} \lambda \ =\ -(D-\sigma h)\epsilon +\frac{\ii}{2}\epsilon^{\mu\nu\rho}\mathcal{F}_{\mu\nu}\gamma_\rho\epsilon 
+ \left(\ii {D}_\mu \sigma - V_\mu \sigma\right)\gamma^\mu\epsilon~.
\eea
Taking the contraction of $\mathcal{Q}\lambda$ with $\epsilon^\dagger$ and 
${\epsilon^c}^\dagger$ separately we obtain two equations. Using the frame 
(\ref{frame}), the first gives
\bea\label{Bog3}
{D}_3\sigma + \mathcal{F}_{12} &=& -\ii(D+ \sigma h)~,
\eea
where we have used the (gauge-dependent) formula (\ref{V3}), while the second gives 
\bea\label{Bog12}
{D}_1\sigma - V_2 \sigma + \mathcal{F}_{23} &=& -\ii \left({D}_2\sigma + V_1\sigma + 
\mathcal{F}_{31}\right)~.
\eea
Notice that both sides of (\ref{Bog3}) and (\ref{Bog12}) must separately be zero, due to the reality conditions on the fields. We conclude that
\bea\label{localizeD}
D+ \sigma h &=& 0~,
\eea
together with the modified Bogomol'nyi equation
\bea\label{Bog}
\Omega^{-1}{D}_\mu\left(\Omega\sigma\right) + \frac{1}{2}\epsilon_\mu^{\ \, \nu \rho}\mathcal{F}_{\nu\rho} &=& 0~.
\eea
Here we have used that $\partial_3\Omega=0$, which follows since $\partial_\psi\Omega=0$. 
In order to proceed with the usual argument, we note that (\ref{Bog}) implies 
that $\Omega\sigma$ is harmonic with respect to the \emph{conformally related
metric} $\dd\hat{s}^2\equiv \Omega^{-2}\diff s^2$. Notice that in this conformal metric one has effectively set $\Omega\equiv 1$ (after relabelling $c\rightarrow \Omega c$), {\it cf.} (\ref{metric}).
We may then rewrite (\ref{Bog}) as 
\bea\label{conformalBog}
D_\mu\left(\Omega\sigma\right) &=& - \left(\hat{*}\, \mathcal{F}\right)_\mu~,
\eea
where $\hat{*}$ denotes the Hodge star operator for the metric $\dd\hat{s}^2$.
It is crucial here 
that $\Omega$ is nowhere zero, so that this conformal metric on $S^3$ is 
also smooth. 
 Then since $M_3\cong S^3$ is compact and simply-connected we conclude that 
\bea\label{localizevector}
\Omega\sigma &=& \mathrm{constant} \ \equiv \ \sigma_0 ~,\nn\\
\mathcal{A}&=& 0~,
\eea
where the second equation follows by substituting back into (\ref{Bog}) and using 
that a flat connection on $S^3$ is trivial. Thus the localization locus in the 
vector multiplet sector is a straightforward modification of the cases 
studied so far in the literature, where in particular $\Omega$ is constant. 

In reaching this conclusion the property of $M_3$ we are using is that $\pi_1 (S^3)$ is trivial, so that there are no non-trivial flat connections. However we may easily extend to the case where $\pi_1 ( M_3 ) \cong\Gamma$, where $\Gamma$ is a finite group. In this case the flat connections are in one-to-one correspondence with homomorphisms from $\Gamma \rightarrow G$, up to conjugacy. One then sums over these flat connections in the localized path integral, in addition to integrating over $\sigma_0$. It is straightforward to extend 
our results, including the one-loop determinants, to this case.

\subsection{Matter multiplet}

An ${\cal N}=2$ chiral multiplet consists of a complex scalar $\phi$, a spinor $\psi$ and an auxiliary field $F$. We take this to be in an arbitrary representation ${\cal R}$ of the gauge group and assign arbitrary R-charge $\rc$. 
The supersymmetry transformations take the form
\begin{align}\label{SUSYmatter}
\mathcal{Q}\phi^\dagger \ =& \ -\psi^\dagger\epsilon~, & \mathcal{Q}\phi \ \ =& \ 0~,\nn\\
\mathcal{Q}\psi \ =& \ \left(\ii\sla{D} \phi + \ii \sigma\phi + \rc h \phi\right)\epsilon~, & \mathcal{Q}\psi^\dagger \ =& \ (\epsilon^c)^\dagger F^\dagger~,\nn\\
\mathcal{Q}F \ =& \ (\epsilon^c)^\dagger\left(\ii\sla{D}\psi - \ii \sigma\psi -\frac{1}{2} \gamma^\mu V_\mu \psi -(\rc-\tfrac{1}{2}) h \psi   \right)~, & \mathcal{Q}F^\dagger \ =& \ 0~,
\end{align}
where the covariant derivatives are
\begin{eqnarray}
D_\mu \phi&=& \left[\nabla_\mu + \ii \rc\left(A_\mu -\tfrac{1}{2} V_\mu \right) + \ii\mathcal{A}_\mu\right] \phi~, \nn\\
D_\mu \psi&=& \left[\nabla_\mu + \ii (\rc-1)\left(A_\mu -\tfrac{1}{2} V_\mu \right) + \ii\mathcal{A}_\mu \right] \psi~.
\end{eqnarray}
Here $\mathcal{A}$ is understood to act in the appropriate representation and $(\sigma \phi)_A=\sigma^i(T_i)^B{}_A \phi_B$, where $A,B$ are indices in ${\cal R}$, $i$ is an index of the Lie algebra and $(T_i)^B{}_A $ are the generators of the gauge group in the representation ${\cal R}$. 

Supersymmetry localizes all these fields to zero, in particular implying that the matter Lagrangian does not contribute to the path integral. To see this, remember that localization requires $\mathcal{Q} \psi^\dagger = 0$ and $\mathcal{Q} \psi =0$. The supersymmetry variation $\mathcal{Q}\psi^\dagger=0$ implies that the F-term $F^\dagger=0$ on the localization 
locus, while $\mathcal{Q} \psi =0$ reads\footnote{It is straightforward to add a real mass for this multiplet by shifting $\sigma$.}
\bea\label{GeneralMatterLocalEqn}
0 &=& - \ii \mathcal{Q}\psi \ = \ (\sla{D}\phi) \epsilon + (\sigma - \ii \rc h ) \phi  \epsilon ~,
\eea
Here the dynamical gauge field $\mathcal{A}_\mu$ vanishes in the localized background, and thus 
is not present in the covariant derivative.
Contracting $\mathcal{Q} \psi$ with $\epsilon^\dagger$ and ${\epsilon^c}^\dagger$ respectively leads to
\bea\label{MatterAbel1}
D_3 \phi \ &=& \ ( - \sigma + \ii \rc h ) \phi ~,
\eea
and
\bea\label{MatterAbel2}
( D_1 + \ii D_2 ) \phi \ &=& \ 0 ~.
\eea
If we consider \eqref{MatterAbel1}, expand $D_3$ and use $\partial_3 = \Omega^{-1} \partial_\psi$ then we find 
\bea
\partial_\psi \phi \ &=& \ \left[ - \Omega \sigma + \ii \Omega \rc \left(h - \tfrac{1}{2} V_3 + A_3 \right) \right] \phi \, .
\eea
In the gauge $V_3 = 2 h$ and using $\sigma = \Omega^{-1} \sigma_0$ this simplifies to 
\bea
\partial_\psi \phi \ &=& \ ( - \sigma_0 + \ii \Omega \rc A_3 ) \phi \, .
\eea
Solving gives\footnote{Here we are using the fact that $A_3$ is independent of $\psi$.}
\bea\label{phisol}
\phi \ &=& \ \ex^{( - \sigma_0 + \ii \Omega \rc A_3 ) \psi} f(z,\bar{z}) \, .
\eea
In a similar manner \eqref{MatterAbel2} can be seen to be
\bea\label{dbar}
\partial_{\bar{z}} \phi \ &=& \ a_{\bar{z}} ( - \sigma_0 + \ii \Omega \rc A_3 ) \phi + \ii c \rc \left( A_{\bar{z}} - \tfrac{1}{2} V_{\bar{z}} \right) \phi ~.
\eea

There are now two cases to consider. Either all the orbits of $K$ are circles, and one has a $U(1)$ 
isometry, or else the generic orbit of $K$ is non-compact. In the first case,  we see immediately from 
(\ref{phisol}) that the solution $\phi$ is not single-valued on the Coulomb branch where $\sigma_0\neq 0$, unless 
$\phi$ is identically zero. In the second case, since the isometry group of a compact manifold is compact, 
we must have at least $U(1)\times U(1)$ symmetry. Unless $\phi$ is identically zero, then 
using (\ref{phisol}) the solution is now unbounded on a dense subset of a copy of the torus inside $S^3$, 
and in particular the solution cannot be continuous.


\section{The partition function}

\subsection{Classical action}\label{SecClassicalAction}

There are two contributions to the localized partition function: the classical action evaluated on the 
localization locus, and the one-loop determinant around the background. In this section 
we evaluate the classical action, showing that it depends only on the 
Killing Reeb vector field $K$.

On the localization locus the only contribution to the classical action for a Chern-Simons-matter theory 
comes from the Chern-Simons Lagrangian (\ref{CSaction}), namely
\bea\label{barry}
S_{{CS}} &=& \int_{M_3}\frac{k}{4\pi}2\ii D\sigma~.
\eea
This is because all other fields (fermions, $\mathcal{A}$, $\phi$, and the F-term) are zero.
Substituting the localization equations (\ref{localizeD}) into (\ref{barry}) we find
\bea
S_{{CS}} &=& -\frac{\ii k}{2\pi}\mathrm{Tr}(\sigma_0^2)\int_{M_3} \frac{h}{\Omega^2}\sqrt{\det g}\, \dd x^3~.
\eea
At first it looks hopeless to evaluate the integral over $M_3$, because the functions are not explicitly known 
in general. However, we may here invoke the existence of an almost contact structure. 
Let us define the following form
\bea
\eta & \equiv & \frac{1}{\Omega}e^3\ = \ \diff\psi+a~.
\eea
In particular this is nowhere zero, since $e^3$ is an almost contact form and the function $\Omega$ is nowhere 
zero \cite{Closset:2012ru}. Moreover, the  Killing vector field $K=\partial_\psi$ is its
Reeb vector field, {\it i.e.}\ we have the equations
\bea
K \, \lrcorner \, \eta &=& 1~, \qquad K \, \lrcorner \, \dd\eta \ = \ 0~.
\eea
Then we compute
\bea
* (\eta\wedge \dd\eta) &=& \frac{1}{\Omega^2}*(e^3\wedge \dd e^3) \ =\ \frac{2h}{\Omega^2}~,
\eea
where we used (\ref{Vh}). Thus
\bea\label{ACvol}
S_{{CS}} &=& -\frac{\ii k}{4\pi}\mathrm{Tr}(\sigma_0^2)\int_{M_3}\eta\wedge \dd\eta~,
\eea
so the integral is precisely the ``almost contact volume'' of $M_3$. 
In general the function $h$ need not be nowhere zero, so that $\eta$ is 
not necessarily a contact form. Nevertheless, the volume appearing in (\ref{ACvol})
depends only on the Reeb vector field $K$, {\it i.e.}\ any two almost contact forms related by continuous deformation with the same Reeb vector field have the same volume. This is 
proven for contact forms in appendix B of \cite{Gabella:2010cy}. 
In fact specializing to $M_3\cong S^3$ 
with a toric contact structure, so that we have $U(1)\times U(1)$ symmetry, we may compute 
this volume explicitly using the Duistermaat-Heckman localization formula in \cite{Martelli:2006yb}. 
If we realize $M_3\cong S^3$ as the contact boundary of $\C^2$ with standard 
symplectic structure, then we may write
\bea\label{K}
K &=& b_1\partial_{\varphi_1}+b_2\partial_{\varphi_2}~,
\eea
where $\varphi_1$, $\varphi_2$ are standard $2\pi$-period coordinates on $U(1)\times U(1)$, and 
since necessarily $b_1,b_2\neq 0$, without loss of generality we may choose orientations so that
$b_1,b_2>0$. 
More precisely, any such toric contact structure on $S^3$ is  induced on the hypersurface $\{\rho=1\}$ of the standard symplectic structure on $\C^2$, namely $\omega =\rho_1\diff \rho_1\wedge \diff\varphi_1 + \rho_2\diff \rho_2\wedge \diff\varphi_2$ with radial coordinate $\rho^2 =b_1\rho_1^2+b_2\rho_2^2$, where the complex coordinates are $z_i=\rho_i\ex^{\ii\varphi_i}$, $i=1,2$. 
The Duistermaat-Heckman theorem then gives
\bea\label{contactvol}
\int_{S^3}\eta\wedge\diff\eta &=& \frac{(2\pi)^2}{b_1b_2}~.
\eea
This localizes the contact volume of $M_3\cong S^3=\partial\C^2$ to the origin of $\C^2$, which is the fixed point 
set of $K$. In the more general case of an almost contact, but not contact, structure, 
the final formula (\ref{contactvol}) still holds since we may always deform 
an almost contact form to a contact form via $\eta\rightarrow \eta+\lambda$, where $\lambda$ is 
a basic one-form for the foliation defined by $K$; this deformation then manifestly 
leaves the volume in (\ref{ACvol}) invariant.
Putting everything together, we get the elegant formula
\bea
S_{CS} &=& -\frac{\ii\pi k}{b_1b_2}\mathrm{Tr}(\sigma_0^2)~,
\eea 
which is the classical contribution to the partition function (\ref{Z}) presented in the introduction.


\subsection{One-loop determinants}\label{SecOneLoop}

We next turn to the one-loop determinants.
The background fields are all zero, except 
$D=-\sigma h$ where $\sigma = \sigma_0/\Omega$ with $\sigma_0$ constant. 
An important fact is that the linearized fluctuations in the matter and vector multiplets
decouple, so that one can compute the one-loop determinants in these sectors independently. In what follows all fields should be understood to be linearized fluctuations
around their background values. 

\subsubsection{Vector multiplet}

In order to compute the one-loop determinant in the vector multiplet sector, we first 
linearize the supersymmetry transformations (\ref{SUSYvector}) around the background. 
Since the only non-zero background field is $\sigma_0$, this is particularly straightforward:
\begin{align}\label{SUSYvectorlin}
\mathcal{Q}\mathcal{A}_\mu \ =& \ -\frac{\ii}{2}\lambda^\dagger\gamma_\mu\epsilon~, & \mathcal{Q}\sigma \ \ =& \ \frac{1}{2}\lambda^\dagger\epsilon~,\nn\\
\mathcal{Q}\lambda \ =& \ \Big[-(D-\sigma h)+\frac{\ii}{2}\epsilon^{\mu\nu\rho}\gamma_\rho\mathcal{F}_{\mu\nu} + \left(\ii{\partial}_\mu\sigma + [\mathcal{A}_\mu,\frac{\sigma_0}{\Omega}]- V_\mu \sigma\right)\gamma^\mu\Big]\epsilon~, & \mathcal{Q}\lambda^\dagger \ =& \ 0~,\nn\\
\mathcal{Q}D \ =& \ \frac{\ii}{2}\partial_\mu(\lambda^\dagger\gamma^\mu\epsilon) + \frac{1}{2}V_\mu\lambda^\dagger\gamma^\mu\epsilon 
- \frac{1}{2}h\lambda^\dagger\epsilon - \frac{\ii}{2}[\lambda^\dagger\epsilon,\sigma]~,
\end{align}
where now all fields are understood to be linearized fluctuations around the background, with 
$\sigma_0$ fixed.  
The $\mathcal{Q}$-exact localizing term is given by
\bea
\mathcal{L}^{\mathrm{loc}}_{\mathrm{vector}} &=& \mathrm{Tr}\, \left[\mathcal{Q}\left((\mathcal{Q}\lambda)^\dagger \cdot \lambda\right)\right]~,\nn\\
&=& \mathrm{Tr}\, \left[(\mathcal{Q}\lambda)^\dagger \cdot (\mathcal{Q}\lambda) + \mathcal{Q}(\mathcal{Q}\lambda)^\dagger \cdot \lambda\right]~,\nn\\
&=& \mathcal{L}^{\mathrm{loc}}_{\mathrm{B}} + \mathcal{L}^{\mathrm{loc}}_\lambda~.
\eea

We begin with the bosonic part $\mathcal{L}^{\mathrm{loc}}_{\mathrm{B}}$. 
This is by construction positive semi-definite, and is zero if and only if $\mathcal{Q}\lambda=0$, 
which are the localization equations of section \ref{SecSusyAndLocalization}. Recall 
that the latter were more transparent 
when expressed in terms of the conformally related metric $\diff \hat{s}^2 \equiv 
\Omega^{-2}\diff s^2$. In particular, the Bogomol'nyi equation (\ref{Bog}) takes 
the standard form (\ref{conformalBog}) when expressed in the conformally related metric. 
As we shall see, the same will be true for the one-loop determinant. 

Given the above comments on positive semi-definiteness of  $\mathcal{L}^{\mathrm{loc}}_{\mathrm{B}}$, 
it is perhaps not surprising to find that, after a simple calculation, one obtains 
\bea
 \mathcal{L}^{\mathrm{loc}}_{\mathrm{B}} &=& \mathrm{Tr}\, \left\{\Omega\left[ \Omega^{-1}\left(\partial_\mu(\Omega\sigma)-\ii [\mathcal{A}_\mu,\sigma_0]\right)+ *\mathcal{F}_\mu 
 \right]^2 + \Omega(D+\sigma h)^2\right\}~,
\eea
where the factors of $\Omega$ arise from $\epsilon^\dagger\epsilon = \Omega$. 
In turn, we may write the corresponding action as a sum of squares using the 
conformally related metric. More specifically, integrating by parts and using 
the Bianchi identity for $\mathcal{F}=\diff\mathcal{A}$, cyclicity of the trace, and 
the gauge-fixing condition $\hat{\nabla}^\mu\mathcal{A}_\mu=0$ we introduce below,  
 one obtains the action
\bea\label{Svec}
\mathcal{S}^{\mathrm{loc}}_{\mathrm{B}} \ =\  \int_{M_3} \sqrt{\det \hat{g}}\, \mathrm{Tr}\, \Big\{
\frac{1}{2}\mathcal{F}_{\mu\nu}^2 - [\mathcal{A}_\mu,\sigma_0]^2+ (\partial_\mu\hat{\sigma})^2 
+ \Omega^4\left(D+\frac{\hat\sigma h}{\Omega}\right)^2~\Big\}~.
\eea
Here the Riemannian measure $\sqrt{\det g} = \Omega^3\sqrt{\det \hat{g}}$, 
all indices are raised using the conformally related metric, and we have defined 
$\hat{\sigma}\equiv \Omega\sigma$.  

This has now essentially reduced 
the bosonic one-loop determinant to the same form as on the round sphere 
studied in \cite{Kapustin:2009kz}, except that the one-loop operators 
appearing are for the conformally related metric. In particular, 
we will add gauge-fixing terms and ghosts appropriate 
to fix the gauge $\hat{\nabla}^\mu \mathcal{A}_\mu=0$, so
\bea
\mathcal{S}_{\mathrm{gauge-fixing}} &=& \int_{M_3}\sqrt{\det \hat{g}}\, \mathrm{Tr}\, \left\{
\bar{c}\hat{\nabla}^\mu\hat{\nabla}_\mu c + b\hat{\nabla}^\mu\mathcal{A}_\mu\right\}.
\eea

As in \cite{Kapustin:2009kz}, we may now immediately do the path integral 
over some of these fields. The integral over $D$ simply sets 
the last term in (\ref{Svec}) to zero. The integral over $\hat{\sigma}$ 
introduces $\det (-\hat{\nabla}^2)^{-1/2}$. The integral over $b$ 
enforces the gauge-fixing condition that $\hat{\nabla}^\mu\mathcal{A}_\mu=0$.
Writing $\mathcal{A}_\mu$ as
\bea
\mathcal{A}_\mu &=& \partial_\mu \varphi + B_\mu~,
\eea
where $\hat{\nabla}^\mu B_\mu=0$, the path integral over $\varphi$ 
and $\hat{\sigma}$ then precisely cancels the 
contribution from the ghosts $c, \bar{c}$. 
The final gauge-fixed action is then simply
\bea\label{gfs}
\mathcal{S}^{\mathrm{loc}}_{\mathrm{B}} &=& \int_{M_3} \sqrt{\det \hat{g}}\, \mathrm{Tr}\, \left\{-
B^\mu \hat{\Delta}B_\mu - [B_\mu,\sigma_0]^2\right\}~,
\eea
where $\hat{\Delta}$ is the vector Laplacian for the metric $\diff \hat{s}^2$.  
On a general three-manifold one cannot hope to compute the 
spectrum of this in closed form (unlike for the round sphere in \cite{Kapustin:2009kz}). However,
most of the eigenvalues of this operator will in fact cancel against the 
eigenvalues of the Dirac operator in the fermionic sector, to which we now turn.

The fermionic localizing term may be written
\bea
 \mathcal{L}^{\mathrm{loc}}_\lambda &=& \mathrm{Tr}\, \left\{\lambda^\dagger \Delta_\lambda \lambda\right\}~,
\eea
where after a lengthy computation one finds the linearized Dirac operator
\bea\label{Paul}
\Delta_\lambda \lambda &=& \ii\Omega \sla{D}\lambda + \ii [\sigma_0,\lambda] 
+\frac{1}{2} \Omega h \lambda - \frac{1}{2}\Omega V_3\lambda 
-\Omega V_\mu\gamma^\mu \lambda + \frac{\ii}{2}(\partial_\mu\Omega)\gamma^\mu\lambda~.
\eea
Here the covariant derivative is
\bea
D_\mu \lambda &=& \nabla_\mu \lambda - \ii (A_{\mu}-\tfrac{1}{2}V_\mu)\lambda~.
\eea
In particular, notice that (\ref{Paul}) is invariant under the $\kappa$ shift gauge symmetry (\ref{shifts}).
Despite the complicated form of (\ref{Paul}), in fact the eigenmodes of this operator 
precisely pair with the eigenmodes appearing in the 
bosonic one-loop action.

\subsubsection*{Pairing of modes}

Suppose first that $\Lambda$ is an eigenmode of $\Delta_\lambda$, with eigenvalue
$M$. That is, $\Delta_\lambda \Lambda = M \Lambda$. Then consider 
\bea\label{Bmode}
B_\mu &\equiv & \partial_\mu (\Omega\epsilon^\dagger\Lambda) + (\ii M  + \alpha(\sigma_0))\epsilon^\dagger 
\gamma_\mu \Lambda~,
\eea
where $\alpha$ denotes the roots of the Lie algebra of the gauge group. By direct computation 
one finds that $B_\mu$ satisfies
\bea\label{firstorder}
\hat{*}\diff B &=& -(M-\ii \alpha(\sigma_0))B~,
\eea
where $\hat{*}$ denotes the Hodge star operator for $\diff \hat{s}^2$. Notice this 
equation immediately implies that $\hat{\nabla}^\mu B_\mu=0$, so that the gauge-field 
mode satisfies the gauge-fixing condition.
Conversely, one can similarly show that if $B_\mu$ satisfies (\ref{firstorder}) 
then
\bea\label{Lambdamode}
\Lambda &\equiv & \gamma^\mu B_\mu \epsilon~, 
\eea
is an eigenvalue of $\Delta_\lambda$ with eigenvalue
$M$. 

The pairing of modes under these first order operators is therefore rather simple.
To apply these results to the one-loop determinants, we observe that 
the quadratic operator appearing in (\ref{gfs}) is
\bea
\Delta_{\mathrm{vec}} &=& \hat{\Delta} +\alpha(\sigma_0)^2~.
\eea
It is then a somewhat standard result to show that the 
transverse eigenmodes of this operator are in one-to-one 
correspondence with modes satisfying the first order equation (\ref{firstorder}). 
More precisely, complex modes $B$ obeying
\bea
\Delta_{\mathrm{vec}} B &=& (M^2 - 2\ii M \alpha(\sigma_0))B~,
\eea
are in one-to-one correspondence with solutions to
\bea
\pm \hat{*} \diff B &=& -(M-\ii \alpha(\sigma_0))B~.
\eea
Using this fact, these paired modes all cancel in the one-loop determinant.

\subsubsection*{One-loop determinant}

Using the results of the previous section we see that  it is only 
the unpaired modes which contribute to the partition function. That is, 
modes for which either (\ref{Bmode}) or (\ref{Lambdamode}) are identically zero. 
We consider the two cases in turn:
\begin{enumerate}
\item  The first unpaired modes are spinor eigenmodes for the Dirac-type operator that pair with identically zero vector eigenmodes. Since $\epsilon$, $\epsilon^c$ span the spinor space, we may write
\bea\label{Lambdaansatz}
\Lambda &=& \epsilon \Phi_0 + \epsilon^c \Phi_2~,
\eea
where $\Phi_0$ and $\Phi_2$ have R-charge 0 and 2, respectively.  
 In order to compute the eigenvalues $M$ in closed form, it is convenient to focus on the case with $U(1) \times U(1)$ isometry. In this case we can write the Killing vector in the form (\ref{Ku1u1}). We may then also 
 expand in Fourier modes
 \bea\label{vecmodes}
 \Phi_0 &=& f_{0,m,n}(\theta)\ex^{-\ii(m\varphi_1+n\varphi_2)}~, \quad  \ \Phi_2 \ =\  f_{2,m,n}(\theta)\ex^{-\ii[(m-1)\varphi_1+(n-1)\varphi_2]}~,
 \eea
 and $\theta$ is any choice of third coordinate.
Here we have used that the dependence of the Killing spinor on $\varphi_1$ and $\varphi_2$ is fixed by the requirement of having a smooth spinor on the three-manifold:
 \begin{equation}\label{spinorphase}
 \epsilon \  = \ \ex^{{\ii}(\varphi_1+\varphi_2)/2} \hat \epsilon  ~,
 \end{equation}
 where $\mathcal{L}_{\partial_{\varphi_1}}\hat \epsilon=\mathcal{L}_{\partial_{\varphi_2}}\hat \epsilon=0$. In particular, the phases  in (\ref{spinorphase}) are then uniquely fixed by requiring the spinor to be smooth at the poles, 
precisely as in the analysis in \cite{Martelli:2012sz, Martelli:2013aqa}.

The eigenvalues $M$ are then immediately determined by substituting (\ref{Lambdaansatz}), (\ref{vecmodes})  
into (\ref{Bmode}), from the 3-component. Using $\partial_3=\frac{1}{\Omega}\partial_\psi$ 
the factors of $\Omega$ cancel, giving
\begin{equation}
\label{eigenv}
M \ = \ m b_1 + n b_2 +\ii \alpha(\sigma_0)
\end{equation}
where $\alpha$ runs over the roots of the gauge group. Normalizability 
of these modes, which is determined by analyzing the first order linear differential 
equation in $\theta$ arising from the remaining 1 and 2-components, requires 
$m,n$ to be non-negative integers, but not both zero {\it i.e.} the mode $m=n=0$ is not a normalizable spinor eigenmode. Again, these statements follow from regularity at the poles.

\item The second class 
are vector eigenmodes that pair with identically zero spinor eigenmodes via (\ref{Lambdamode}). 
This means $\gamma^\mu B_\mu\epsilon=0$, which immediately implies $B_3=0$ and $B_1+\ii B_2=0$. 
We may thus write $B=B_1e^1+B_2e^2$ and compute $\diff B$ appearing in the 
bosonic equation (\ref{firstorder}) using the frame definitions $e^i=c(x_1,x_2)\diff x^i$, $i=1,2$. 
Specifically, expanding in Fourier modes
\bea
B_1 &=& b_{m,n}(\theta)\ex^{-\ii(m\varphi_1+n\varphi_2)}~,
\eea
the factors of $\Omega$ again cancel and one finds precisely the eigenvalues (\ref{eigenv}), but now normalizability requires $m,n \leq -1$. 
\end{enumerate}

The first class contributes to the numerator while the second class contributes to the denominator in the one-loop determinant of the vector multiplet. Putting everything together we get
\begin{equation}\label{vectorprod}
\prod_{\alpha \in \Delta} \frac{1}{\ii \alpha(\sigma_0)} \prod_{m,n \geq 0} \frac{b_1 m + b_2 n  + \ii \alpha(\sigma_0)}{-(m+1)b_1 -(n+1)b_2 + \ii \alpha(\sigma_0)}~,
\end{equation}
where $\Delta$ denotes the set of roots. The infinite product may be regularized using zeta function regularization, 
the steps involved being a trivial modification of those appearing in appendix C of \cite{Martelli:2011fu}. 
In this way we may write this expression as a product of $\sinh$ functions, which when combined 
with the Vandermonde determinant leads precisely to the result quoted in the introduction. 

\subsubsection{Matter multiplet}

The linearized supersymmetry transformations in the matter sector are
\begin{align}\label{SUSY}
\mathcal{Q}\phi^\dagger \ =& \ -\psi^\dagger\epsilon~, & \mathcal{Q}\phi \ \ =& \ 0~,\nn\\
\mathcal{Q}\psi \ =& \ \left(\ii\sla{D} \phi + \ii \frac{\sigma_0}{\Omega}\phi + \rc h \phi\right)\epsilon~, & \mathcal{Q}\psi^\dagger \ =& \ (\epsilon^c)^\dagger F^\dagger~,\nn\\
\mathcal{Q}F \ =& \ (\epsilon^c)^\dagger\left(\ii\sla{D}\psi - \ii \frac{\sigma_0}{\Omega}\psi -\frac{1}{2} \gamma^\mu V_\mu \psi -(\rc-\tfrac{1}{2}) h \psi   \right)~, & \mathcal{Q}F^\dagger \ =& \ 0~,
\end{align}
where the covariant derivatives are
\begin{eqnarray}\label{covD}
D_\mu \phi&=& \left[\nabla_\mu + \ii \rc\left(A_\mu -\tfrac{1}{2} V_\mu \right)\right] \phi~, \nn\\
D_\mu \psi&=& \left[\nabla_\mu + \ii (\rc-1)\left(A_\mu -\tfrac{1}{2} V_\mu \right) \right] \psi~.
\end{eqnarray}
Notice 
that the fluctuation of $\sigma$ around $\sigma_0/\Omega$ does not appear, 
as it multiplies the background value $\phi=\phi_0=0$. A similar comment 
applies to the fluctuation of the gauge field $\mathcal{A}_\mu$ around its background value of 
zero.

We begin by rewriting the supersymmetry transformations (\ref{SUSY}) in terms of 
 operators that map functions to spinors (in the representation $\mathcal{R}$). 
 This will turn out to be a particularly convenient way to compute the wave operators 
 that appear in the one-loop determinant, and will make the pairing of modes under 
 supersymmetry manifest. 

We first define operators $S_1,S_2$, mapping functions to spinors, via
\bea
S_1\Phi &=& \Phi\epsilon~, \nn\\
S_2\Phi &=& \left(\ii\sla{D}\Phi + \ii \frac{\sigma_0}{\Omega}\Phi+ \rc h \Phi \right)\epsilon~.
\eea
 We define the obvious inner products on complex functions and spinors as
\bea
\langle \Phi_1,\Phi_2\rangle &=& \int_{M_3}\sqrt{\det g}\, \Phi_1^\dagger\Phi_2~, \qquad \langle \Psi_1,\Psi_2\rangle \ = \  \int_{M_3} \sqrt{\det g}\, \Psi_1^\dagger\Psi_2~,
\eea
and then compute the adjoint operators with respect to these:
\bea
S_1^*\Psi &=& \epsilon^\dagger \Psi~,\nn\\
S_2^*\Psi &=& \epsilon^\dagger\left( \ii \sla{D}  - \ii\frac{\sigma_0}{\Omega} + \frac{1}{2} \gamma^\mu V_\mu + (\rc-\tfrac{3}{2})h \right) \Psi~,
\eea
where we used the useful identities
\bea\label{slaKSE}
\sla{D} \epsilon  & =& + \frac{\ii}{2} \gamma^\mu V_\mu \epsilon-\frac{3}{2}\ii h \epsilon~,\nn\\
\sla{D} \epsilon^c & =& - \frac{\ii}{2} \gamma^\mu V_\mu \epsilon^c -\frac{3}{2}\ii h \epsilon^c~,
\eea
derived from \eqref{KSE}, \eqref{KSEconj}.
We similarly define the conjugate operators as
\bea
S_1^c\Phi &=& \Phi\epsilon^c~, \nn\\
S_2^c\Phi &=&  \left( \ii \sla{D}\Phi + \ii\frac{\sigma_0}{\Omega} \Phi- (\rc-2)h  \Phi \right) \epsilon^c~.
\eea
The adjoints are
\bea
S_1^{c*}\Psi &=& {\epsilon^{c}}^\dagger \Psi~,\nn\\
S_2^{c*}\Psi &=& {\epsilon^{c}}^\dagger\left( \ii \sla{D}  - \ii\frac{\sigma_0}{\Omega} - \frac{1}{2} \gamma^\mu V_\mu - (\rc-\tfrac{1}{2})h \right) \Psi~.
\eea

Using these definitions, the linearized supersymmetry variations (\ref{SUSY}) become
\bea\label{SUSYS}
\mathcal{Q}\phi^\dagger &=& -(S_1^*\psi)^\dagger~, \, \ \ \qquad \qquad \mathcal{Q}\phi \ = \ 0~,\nn\\
\mathcal{Q}\psi &=& S_2\phi~, \, \ \qquad \qquad \qquad \mathcal{Q}\psi^\dagger \ = \ (S_1^cF)^\dagger~,\nn\\
\mathcal{Q}F &=& S_2^{c*}\psi~, \qquad\qquad \qquad \mathcal{Q}F^\dagger \ = \ 0~.
\eea
Here we note that the covariant derivative appearing in $S_2^c$ is
\bea
D_\mu &=& \nabla_\mu + \ii (\rc-2)\left(A_\mu -\tfrac{1}{2} V_\mu \right)~,
\eea
relevant for a field of R-charge $\rc-2$.

We may now write the $\mathcal{Q}$-exact  matter localizing action in terms of the operators introduced above. 
We define
\bea
{\cal L}_\mathrm{matter}^\mathrm{loc} & = &  \mathcal{Q} \left((\mathcal{Q}\psi)^\dagger \cdot \psi + \psi^\dagger \cdot (\mathcal{Q}\psi^\dagger)^\dagger \right) \nn\\
& = &  (\mathcal{Q}\psi)^\dagger \cdot (\mathcal{Q} \psi)  + (\mathcal{Q}\psi^\dagger ) \cdot (\mathcal{Q}\psi^\dagger)^\dagger 
+  \mathcal{Q}(\mathcal{Q}\psi)^\dagger \cdot \psi - \psi^\dagger \cdot \mathcal{Q} (\mathcal{Q}\psi^\dagger)^\dagger \nn\\
 & \equiv & {\cal L}_\mathrm{\phi}^\mathrm{loc} + {\cal L}_\mathrm{\psi}^\mathrm{loc}~,
\eea
where we used that $\mathcal{Q}$ anti-commutes with $\mathcal{Q}$, $\psi$, $\psi^\dagger$, and 
an overall trace is implicit in these formulae.
The bosonic part ${\cal L}_\mathrm{\phi}^\mathrm{loc}$ contains the F-term
 $(\mathcal{Q}\psi^\dag )\cdot (\mathcal{Q}\psi^\dag)^\dag = F^\dag F$ which does not contribute to the wave operator, together with the term
\bea
\int_{M_3}\sqrt{\det g}\, (\mathcal{Q}\psi)^\dagger \cdot (\mathcal{Q}\psi) &=& \langle S_2\phi, S_2\phi\rangle \ = \ \langle \phi , S_2^*S_2\phi\rangle \ \equiv \ \langle \phi, \Delta_\phi\phi\rangle~,
\eea
which immediately allows us to read off the quadratic wave operator
\bea
\Delta_\phi &=& S_2^* S_2 ~.
\eea
A computation shows that this operator takes the explicit form
\bea
\Delta_\phi &=& -\Omega D^\mu D_\mu \Phi  +  r \Omega \epsilon^{\rho\mu\nu}  e^3_\rho (F_{\mu\nu}  -  \tfrac{1}{2}V_{\mu\nu} ) + 2\ii\Omega  V^\mu D_\mu \Phi - \epsilon^{\rho\mu\nu} \Omega e^3_\rho V_\mu D_\nu  \Phi \nn\\ &&
+ 2 \ii h (r-1) D_3 \Phi  + r^2 h^2 \Omega \Phi + \frac{\sigma_0^2}{\Omega}\Phi ~.
\eea
In order to compute the fermionic counterpart, it is helpful 
to first establish some identities. 
We proceed by introducing the projection operators
\bea
P_\pm &\equiv & \frac{1}{2}(1\pm e^3_\mu\gamma^\mu) \ = \ \frac{1}{2}(1\pm \sigma_3)~.
\eea
The Fierz identity, together with $\epsilon^\dagger\epsilon=\Omega$ and $\epsilon^\dagger \gamma_\mu\epsilon = \Omega e^3_\mu$, then gives
\bea\label{Fierzes}
(\epsilon^\dagger\Psi)\epsilon &=& \Omega P_+\Psi~, \qquad \epsilon^\dagger(\Psi^\dagger\epsilon) \ =\ \Omega\Psi^\dagger P_+~,
\eea
and the conjugate version
\bea
(\epsilon^{c\dagger}\Psi)\epsilon^c &=& \Omega P_-\Psi~.
\eea
It is then straightforward to derive
\bea
S_1^*S_1 &=& \Omega~, \qquad  \ S_1S_1^* \ = \ \Omega P_+~,\nn\\
S_1^{c*}S_1^c &=& \Omega~, \qquad S_1^cS_1^{c*} \ = \ \Omega P_-~.
\eea
The fermionic part ${\cal L}_\mathrm{\psi}^\mathrm{loc}$  comprises the two terms
\bea
\mathcal{Q}(\mathcal{Q}\psi)^\dagger \cdot \psi & \equiv &  \psi^\dagger \Delta^{(+)}_\psi \psi~,\nn\\
- \psi^\dag \cdot \mathcal{Q}(\mathcal{Q} \psi^\dag)^\dag & \equiv  &  \psi^\dagger \Delta^{(-)}_\psi \psi  ~.
\eea
We compute
\bea\label{Deltaplus}
-\Delta^{(+)}_\psi \psi &=& \Omega P_+\left[\ii\sla{D}+\frac{1}{2}\gamma^\mu V_\mu - \ii \frac{\sigma_0}{\Omega}
+(\rc-\tfrac{3}{2})h\right]\psi   \ =\ S_1S_2^*\psi~.
\eea
In deriving 
(\ref{Deltaplus}) we have used both of the identities in (\ref{Fierzes}), and 
one needs to integrate by parts. 
We may handle $\Delta^{(-)}_\psi$ in a similar way:
\bea
\mathcal{Q}(\mathcal{Q}\psi^\dagger)^\dagger &=& \mathcal{Q}\left((S_1^cF)^\dagger\right)^\dagger \ = \ \mathcal{Q}(F\epsilon^c) \nn\\
&=& (S_2^{c*}\psi)\epsilon^c \ = \ S_1^cS_2^{c*}\psi\nn\\
&=& \Omega P_-\left[\ii\sla{D}-\frac{1}{2}\gamma^\mu V_\mu-\ii\frac{\sigma_0}{\Omega}-(\rc-\tfrac{1}{2})h\right]\psi~.
\eea
The total Dirac-type operator $\Delta_\psi$ acting on spinorial wave functions in the one-loop determinant is then
\bea
\Delta_\psi &=& \Delta_\psi^{(+)}+\Delta_\psi^{(-)} \ = \ -S_1S_2^*-S_1^cS_2^{c*}~.
\eea
The operator $S_1S_2^*$ and its conjugate version of course map spinors to spinors. 
The operator $S_2S_1^*$ similarly maps spinors to spinors, but is slightly different. 
This appears in the pairing of modes under supersymmetry, so it is useful to compute:
\bea
S_2S_1^* \ =\  \Omega \ii\gamma^\mu P_+\left[D_\mu +\frac{\ii}{2}h\gamma_\mu + \frac{\ii}{2}V_\mu - \frac{1}{2}\epsilon_{\mu\nu\rho}V^\nu\gamma^\rho\right]+ \Omega\left(\ii\frac{\sigma_0}{\Omega}+\rc h\right)P_+~.
\eea
The conjugate version is similarly
\bea
S_2^cS_1^{c*} \ = \  \Omega \ii\gamma^\mu P_-\left[D_\mu +\frac{\ii}{2}h\gamma_\mu - \frac{\ii}{2}V_\mu + \frac{1}{2}\epsilon_{\mu\nu\rho}V^\nu\gamma^\rho\right]+ \Omega\left(\ii\frac{\sigma_0}{\Omega}-(\rc-2)h\right)P_-~.
\eea
Combining these we find the remarkable formula
\bea\label{twodiracs}
 -S_1S_2^*-S_1^cS_2^{c*} \ =\ \Delta_\psi \ = \  -S_2S_1^*-S_2^cS_1^{c*} + 2\ii\sigma_0~.
\eea
Notice that $\Omega$ drops out of the last term, via $\Omega\cdot \frac{\sigma_0}{\Omega}$, with the first 
factor of $\Omega$ arising from the square norm of the Killing spinor. 

To finish these formalities, before moving on to the pairing of modes under supersymmetry, 
we estabilish some orthogonality relations. First, it is trivial to see that $S_1$ and $S_1^c$ are orthogonal; 
that is,
\bea
\langle S_1^*S_1^c\Phi_1,\Phi_2\rangle &=& \langle S_1^c\Phi_1,S_1\Phi_2\rangle \ = \ \langle \Phi_1,S_1^{c*}S_1\Phi_2\rangle \ = \ 0~,
\eea
holds for all functions $\Phi_1$, $\Phi_2$. This follows immediately from ${\epsilon^c}^\dagger \epsilon = 0$. 
We next claim that $S_2$ and $S_2^c$ are also orthogonal. This is more involved.
After a  computation we find
\bea\label{S2orth}
S_2^*S_2^c\Phi &=& -(\rc-2) \left[ \frac{\ii}{2} \epsilon^\dagger \gamma^{\mu \nu} \epsilon^c \left(F_{\mu \nu} -\tfrac{1}{2} V_{\mu \nu}\right) +\ii \epsilon^\dagger \gamma^\mu \epsilon^c (\partial_\mu h -\ii V_\mu h) \right] \Phi~.
\eea
Here we have introduced $V_{\mu \nu} = \nabla_\mu V_\nu - \nabla_\nu V_\mu $. 
That (\ref{S2orth}) is zero follows  from the integrability condition for the Killing spinor equation (\ref{KSE}). 
Taking a covariant derivative of (\ref{KSE}) and skew symmetrizing leads to\footnote{This 
identity was first derived in \cite{Martelli:2013aqa}.} 
\begin{eqnarray}
\nonumber
\left[ \frac{1}{4} R_{\mu \nu \rho \sigma} \gamma^{\rho \sigma} -\ii F_{\mu \nu} +\frac{3}{2} \ii V_{\mu \nu} +\ii \partial_{[\mu}h \gamma_{\nu ]} -\frac{1}{2} h^2 \gamma_{\mu \nu}+h \gamma_{[\mu} V_{\nu ]}  \right. \\
\left. -\ii \nabla_{[ \mu|} V_\rho \gamma_{\nu]} \gamma^\rho +\frac{1}{2} V^\rho V_\rho \gamma_{\mu \nu} -V_\rho \gamma_{[\mu} V_{\nu]}\gamma^\rho \right] \epsilon \ = \ 0~.
\end{eqnarray}
Taking the charge conjugate of this equation, and applying $\epsilon^\dagger \gamma^{\mu \nu}$ on the left, 
half of the terms are zero and we obtain precisely that
\bea
S_2^*S_2^c\Phi &=& 0~,
\eea
holds for all $\Phi$.

\subsubsection*{Pairing of modes}

After this preparation, it is now quite straightforward to compute the pairing of modes 
in the one-loop determinant.
 Let us first suppose that $\Phi$ is a scalar eigenmode satisfying
\bea
\Delta_\phi \Phi & = &  S_2^* S_2 \Phi \ = \ \frac{\mu}{\Omega} \Phi~.
\eea
The factor of $\Omega$ may be compared to a similar factor that appears in the 
vector bosonic operator (\ref{firstorder}) when it is written in terms of the original 
metric. Indeed, one can write $\Delta_\phi=\frac{1}{\Omega}\hat{\Delta}_\phi$, 
where the leading second order derivative operator appearing in 
$\hat{\Delta}_\phi$ is precisely the Laplacian for the conformal metric $\diff\hat{s}^2$.
We then consider the two associated spinors
\bea
\Psi_1 \ \equiv \ \Omega^{-1} S_1 \Phi ~, \qquad \qquad \Psi_2 \ \equiv \ S_2  \Phi~.
\eea
One computes
\bea
- \Delta_\psi \Psi_2  \ = \  S_1 S_2^*S_2 \Phi  + S_1^c S_2^{c*} S_2 \Phi\ =  \  S_1 S_2^*S_2 \Phi  \ = \   \frac{\mu}{\Omega} S_1 \Phi \ =\   \mu \Psi_1 ~,
\eea
where the second equality follows from $S_2$ and $S_2^c$ being orthogonal. Similarly, upon using the identity (\ref{twodiracs}) to rewrite $\Delta_\psi$, we have
\bea
- \Delta_\psi \Psi_1 \ = \  S_2 S_1^* (\Omega^{-1}S_1 \Phi)   +  S_2^c S_1^{c*} (\Omega^{-1}S_1 \Phi) - 2\ii \sigma_0  \Psi_1  ~,
\eea
which using  that $S_1^{c*}S_1=0$  and $S_1^* S_1=\Omega$ reads
\bea
- \Delta_\psi \Psi_1 \ = \  \Psi_2   - 2\ii \sigma_0  \Psi_1  ~.
\eea
Thus, we have computed
\bea
\Delta_\psi
\left(
\begin{array}{c}
 \Psi_1 \\ \Psi_2
\end{array}
\right) & = &
\left(
\begin{array}{cc}
  2\ii\sigma_0  & -1 \\
  - \mu  & 0 \\
\end{array}
\right)
\,
\left(
\begin{array}{c}
 \Psi_1 \\ \Psi_2
\end{array}
\right) ~   \equiv ~\mathscr{M} 
\left(
\begin{array}{c}
 \Psi_1 \\ \Psi_2
\end{array}
\right)  ~,
\eea
and hence the  eigenvalues are
\bea
\lambda_{\pm} & = & \ii  \sigma_0 \pm  \sqrt{ \mu -\sigma_0^2} ~.
\eea
Thus, if $\mu= M (M - 2 \ii  \sigma_0)$, the eigenvalues are
\bea
\lambda_+ \ =\  M ~,\qquad \qquad \lambda_- \ =   \ -M + 2 \ii \sigma_0 ~.
\eea
We then have a pairing between a scalar eigenmode with eigenvalue $\mu=M (M - 2 \ii  \sigma_0) $
and two spinor eigenmodes of $\Delta_\phi $ with eigenvalues $M, \, 2\ii \sigma_0 -   M$.
Any modes which take part in this pairing can be neglected when computing the one-loop determinant.

Conversely, suppose that $\Psi$ is  eigenmode of the Dirac operator $\Delta_\psi$ with eigenvalue $M$, namely
\bea
\Delta_\psi \Psi & = & M \Psi ~,
\label{spinormode}
\eea
and consider the function $\Phi=S_1^*\Psi$. Hitting (\ref{spinormode}) with $S_2^*$, where $\Delta_\psi $ is written as in the right hand side of (\ref{twodiracs}),  we have
\bea
S_2^* S_2 \Phi + S_2^* S_2^c S_1^{c*} \Psi  - 2\ii \sigma_0  S_2^* \Psi  & =  & - M S_2^*\Psi~ .
\eea
Notice that we have commuted $S_2^*$, a differential operator, with $\sigma_0$ as the latter is constant in the background.
Then using orthogonality of $S_2$ and $S_2^c$  this gives
\bea
S_2^* S_2 \Phi  &  = &  \left( -M + 2\ii \sigma_0  \right) S_2^*\Psi~ .
\eea
On the other hand, hitting (\ref{spinormode}) with $S_1^*$,
where $\Delta_\psi $ is  written as in the left hand side of  (\ref{twodiracs}),  and using orthogonality of $S_1$ and $S_1^c$,
 we have
\bea
- M S_1^*\Psi \ = \   S_1^* S_1 S_2^*  \Psi \ =\ \Omega S_2^*  \Psi~.
\eea
Therefore
\bea
\Delta_\phi \Phi \ = \ S_2^* S_2 \Phi \ = \frac{1}{\Omega} M \left( M -  2\ii \sigma_0  \right)\Phi~,
\eea
so that again $\mu= M \left( M -  2\ii \sigma_0  \right)$.

\subsubsection*{One-loop determinant}

The previous section implies that when computing the ratio
$\det \Delta_\psi/\det \Delta_\phi$, almost all the modes cancel. 
The only contribution to this ratio comes from modes where the 
above pairings between spinors and functions degenerates. There 
are two cases:
\begin{enumerate}

\item Suppose we have a spinor mode $\Psi$ with eigenvalue $M$, so 
$\Delta_\psi\Psi = M\Psi$, but that the putative function mode $\Phi=S_1^*\Psi\equiv 0$. 
Then the contribution of $\Psi$ to $\det \Delta_\psi$ is left uncancelled, and 
we must put this mode back in. These are spinor eigenmodes 
sitting in $\ker S_1^* = \image S_1^c$, where this equality follows from the fact that $\epsilon^c$ and $\epsilon$ form an orthogonal basis of the spinor space. 

Thus we must precisely add back the spinor modes 
$\det \Delta_\psi\mid_{\ker S_1^*}$, which contribute to the numerator 
of $\det\Delta_\psi/\det\Delta_\phi$.  It is easy to work these out explicitly:
\bea
M\Psi &=& \Delta_\psi \Psi \ = \ -S_2^cS_1^{c*}\Psi + 2\ii \sigma_0\Psi~,
\label{getdetone}
\eea
where we have used $S_1^*\Psi=0$. Using the fact that $\Psi = S_1^c\Phi$, 
where $\Phi$ has R-charge $(\rc-2)$, and contracting with $\epsilon^{c\dagger}$ (equivalently applying $S_1^{c*}$),
 one gets the scalar equation
 \begin{equation}
M \Phi \ = \ \left(\ii \partial_{\psi} +\ii \sigma_0- \Omega(\rc -2)(A_3 -\tfrac{1}{2} V_3 - h) \right)\Phi 
 \end{equation}
 In order to compute the eigenvalues $M$ in closed form, it is again convenient to focus on the case with $U(1) \times U(1)$ isometry. In this case we can write the Killing vector in the form (\ref{Ku1u1}), where the 
 dependence of the Killing spinor on $\varphi_1$ and $\varphi_2$ is fixed by (\ref{spinorphase}). 
Using the Killing spinor equation (\ref{KSE}), plus the explicit form of $A_3$ given in appendix \ref{AppMoreFormulae}, one computes
\begin{equation}
\label{constantcomb}
\Omega (A_3 -\tfrac{1}{2} V_3 - h) \ =  \ \frac{b_1 + b_2}{2} ~.
 \end{equation}
 Next we can expand in Fourier modes
 \begin{equation}
 \Phi \ = \  f_{m,n}(\theta)\ex^{-\ii (m \varphi_1 + n \varphi_2)}~,
 \end{equation}
 where $\theta$ is a third coordinate. In  a weight space decomposition of the representation ${\cal R}$ the eigenvalues are therefore given by 
 \begin{equation}
 M \ = \ \ii \rho(\sigma_0)+ b_1 m + b_2 n -  (\rc -2) \frac{b_1+b_2}{2}~,
 \end{equation}
 where $\rho$ are the weights of the representation. Finally, $f_{m,n}(\theta)$ satisfies a first order linear differential equation, obtained by applying $\epsilon^\dagger$ to (\ref{getdetone}). Regularity of the solution to this equation at the poles requires $m,n$ to be non-negative integers.
 
 \item On the other hand, if $\Psi_1 =\Omega^{-1}S_1 \Phi$ and $\Psi_2=S_2 \Phi$ are proportional, for a given function mode $\Phi$, then we have overcounted in $\det \Delta_\psi$. In this case $\Psi_2 =S_2 \Phi$ is also in $\image S_1 = \ker S_1^{c *}$. We may contract the following equation
 \begin{equation}
 S_2 \Phi \ = \ M \Omega^{-1} S_1 \Phi~,
\end{equation}
with $\epsilon^\dagger$. The factors of $\Omega$ again all cancel and one obtains the eigenvalues
\begin{equation}
M \ = \ \ii \rho(\sigma_0)+ b_1 m + b_2 n- \rc \frac{b_1+b_2}{2}~.
\end{equation} 
Normalisability of the modes can be inferred from the dependence of the wave function $f_{m,n}(\theta)$ on $m,n$, which now requires $m,n$ to be non-positive integers. 
 \end{enumerate}

Putting everything together, the one-loop determinant is
\begin{equation}
\label{oneloop}
\prod_{m,n \geq 0} \frac{ b_1 m + b_2 n + \ii \rho(\sigma_0) -  (\rc -2) \frac{b_1+b_2}{2}}{ b_1 m + b_2 n - \ii \rho(\sigma_0)+ \rc \frac{b_1+b_2}{2}} \ = \ s_\beta\left[-\frac{\rho(\sigma_0)}{\sqrt{b_1b_2}} - \frac{\ii Q}{2} (\rc-1)\right]~,
\end{equation}
where $\beta\equiv \sqrt{\frac{b_1}{b_2}}$, $Q\equiv \beta+\beta^{-1}$ and the double sine function is defined by
\begin{equation}
s_\beta(z) \ \equiv \ \prod_{m,n \geq 0 } \frac{\beta m+\beta^{-1} n + \frac{Q}{2} -\ii z }{\beta m+\beta^{-1} n + \frac{Q}{2} +\ii z } ~.
\end{equation}
This completes the derivation of the one-loop matter determinant appearing in (\ref{Z}).


\subsection{Generalizations to other three-manifolds}\label{generalSec}

In this section we briefly comment on generalizing the above results to other background geometries. 

First, let us consider three-sphere geometries with only $U(1)$ symmetry. 
Notice that in order to obtain the unpaired eigenvalues in closed form, above we have assumed 
$U(1)\times U(1)$ symmetry. This symmetry immediately follows if $K$ has a non-closed orbit. 
On the other hand, if $K$ generates a $U(1)$ isometry in fact the same form of the partition function
follows. To see this, we note that in this case the action of $K$ defines 
a Seifert fibration of $S^3$, and these are classified. In particular, 
$K$  is contained in the standard $U(1)\times U(1)$ action on $S^3$ so we may write $K=p\partial_{\varphi_1}+q\partial_{\varphi_2}$, 
where $p,q$ are positive integers with $\mathrm{gcd}(p,q)=1$. Here the background fields are necessarily invariant under  
$K$, but not necessarily under $\partial_{\varphi_1}$, $\partial_{\varphi_2}$ separately.  
In this case we may decompose the eigenmodes in terms of $\psi$-dependence 
and dependence on the transverse space, which is a weighted projective space 
$\mathbb{WCP}^1_{[p,q]}$. That is, one writes a mode as $\Phi=\ex^{\ii\gamma\psi}f(z,\bar{z})$, 
for fixed $\gamma$,  where $f$ then becomes  a section of 
a line bundle over this base space. 
The transverse equation 
for this mode is then a $\bar{\partial}$-type operator. For example, in the matter sector one
obtains the $\bar{\partial}$ operator
\bea
{\partial}_{\bar{z}}-\ii \left[cr(A_{\bar{z}}-\tfrac{1}{2}V_{\bar{z}})-\gamma a_{\bar{z}}\right]~.
\eea
In order to  compute the one-loop determinant,
one then simply computes the degeneracy of this 
operator, which is given by an (orbifold) index theorem. For example, 
we may phrase the computation on the round sphere in this language. 
The eigenvalues for a scalar under the Hopf action of $K=\partial_\psi$ are 
integers $n$, which leads to the $\bar{\partial}$ operator for 
$\mathcal{O}(n)$ on the base space $\mathbb{CP}^1$. The index of this 
is $n+1$ (where $n$ is the Chern number of $\mathcal{O}(n)$ and the factor of 1 comes from the 
Todd class) which is then the degeneracy of this eigenvalue. 

These last comments also lead to a more general method for computing the partition function 
on a manifold with arbitrary topology. Since the background geometry is always equipped with a 
nowhere zero Reeb vector field $K$, it follows that $M_3$ admits a Seifert fibration. If the 
orbits of $K$ are all compact, then the leaves of this fibration may be taken to be the orbits of $K=\partial_\psi$. (On the other hand when $K$ has a non-compact orbit we are in the toric case with $U(1)\times U(1)$ symmetry.) In this case, as in the above 
paragraph we may then Fourier decompose all modes entering the one-loop determinants along the $\psi$ 
direction, so that the transverse equation is then a zero mode equation for an appropriate 
$\bar\partial$ operator on the orbit space $M_3/U(1)$, which is in general an orbifold. The computation 
of the one-loop determinants then amounts to computing the index of this operator. In particular, 
this way of viewing the one-loop computation makes it manifest that it is a topological invariant, 
depending only on the Reeb vector field $K=\partial_\psi$.\footnote{Notice that we already 
showed that the classical contribution to the partition function in section \ref{SecClassicalAction} depends 
only on $K$.}

Finally, let us comment more specifically on the case of three-manifolds with finite fundamental group, 
which are then of the form $S^3/\Gamma$, where $\Gamma$ is a finite group. 
If $\Gamma$ acts as a subgroup of the $U(1)\times U(1)$ symmetry, then it is immediate to 
extend our results to this case, provided $\Gamma$ also preserves the Killing spinor \cite{Alday:2012au}. 
The quotient $S^3/\Gamma$ is then a Lens space, and the form of the partition function 
is as in  \cite{Benini:2011nc,Gang:2009wy,Alday:2012au}. In particular, there is a sum over 
flat connections, in addition to the integral over $\sigma_0$, and in the one-loop determinants 
the sums over integers $m$ and $n$ have projection conditions determined by the choice of 
flat connection. For precise details we refer to the aforementioned papers.


\section{Applications}\label{Applications}

\subsection{Large $N$ limit and comparison to gravity duals}\label{LargeNLimit}

The main result of this paper is an explicit expression for the partition function of an $\mathcal{N}=2$ Chern-Simons theory coupled to generic matter:
\bea
Z \ = \ \int \diff\sigma_0\, \ex^{\frac{\ii\pi k}{b_1b_2}\mathrm{Tr}\, \sigma_0^2}\prod_{\alpha\in \Delta_+} 4
\sinh \frac{\pi\sigma_0\alpha}{b_1}\sinh\frac{\pi\sigma_0\alpha}{b_2}\prod_{\rho}s_\beta\left[\frac{\ii Q}{2}(1-\rc)-\frac{\rho(\sigma_0)}{\sqrt{b_1b_2}}\right]~,
\label{mainresult}
\eea
where $\beta=\sqrt{\frac{b_1}{b_2}}$ and $Q=\beta+\beta^{-1}$. This has the general form studied in \cite{Martelli:2011fu}, where the large $N$ limit was computed for a broad class of Chern-Simons-quiver theories using saddle point methods. The final form for the free energy ${\cal F} = -\log Z$ in the large $N$ limit is
\begin{equation}
{\cal F}_\beta  \  = \  \frac{Q^2}{4} {\cal F}_{\beta=1}~,
\end{equation}
with ${\cal F}_{\beta=1}$ the partition function of the round sphere \cite{Jafferis:2010un}, which scales like $N^{3/2}$.

In \cite{Martelli:2013aqa} the authors found holographic duals to a two-parameter family of deformed  three-sphere backgrounds that fit into the class studied in this paper. 
In all cases considered, the holographic free energy in this reference was written as
\begin{equation}
{\cal F}_{\mathrm{gravity}} \ = \ \frac{\pi}{8 G_4} \left(\beta+\beta^{-1}\right)^2~,
\end{equation}
where $G_4$ is the Newton constant in four-dimensional $\mathcal{N}=2$ gauged supergravity, and $\beta = \beta^\mathrm{MP} (a,v)$ was a function of the two real 
parameters with $a$ and $v$ characterizing the geometries. Indeed, verifying the conjecture made in  \cite{Martelli:2013aqa}, that 
 the full localized partition function on manifolds with three-sphere topology is of the form (\ref{mainresult}) for 
 some appropriate definition of $\beta$, was one of the motivations for our work.

Let us show more explicitly how our field theory result for the partition function matches precisely the gravity predictions in \cite{Martelli:2013aqa}.
First of all, note that although the backgrounds considered in \cite{Martelli:2013aqa} contained generically complex fields $A_\mu$ and $V_\mu$, 
in all cases there is a region of the parameter space 
where everything is real. We will compare our field theory results to these real backgrounds, commenting at the end on the more general complex case. 
The key point is that the coordinates $\varphi_1$, $\varphi_2$ used in \cite{Martelli:2013aqa} are precisely the coordinates used 
presently,\footnote{Up to some irrelevant sign changes $\varphi_i \to - \varphi_i$.} and in particular the Killing spinors depend on these 
exactly as in (\ref{spinorphase}). 
The metric and other background fields were not written explicitly in these coordinates, but given our findings, this is not necessary. 
All we need is to write the Killing Reeb vector field, which is simple to obtain from the dual almost contact one-form given in \cite{Martelli:2013aqa}
\bea
e^3 \,  = \,  \left|\Big(p_3^2-p^2 + \frac{\mathcal{P}(p)}{p^2+\sqrt{\alpha}}\Big) \frac{2}{\mathcal{P}'(p_3)}\right| \diff\varphi_1 + \left|\Big(p_4^2-p^2 + \frac{\mathcal{P}(p)}{p^2+\sqrt{\alpha}}
\Big) \frac{2}{\mathcal{P}'(p_4)}\right| \diff\varphi_2 ~.
\eea
Here $\mathcal{P}(p)$ is a quartic polynomial in the variable $p$, with (at least) two real roots $p_3,p_4$ parameterizing the background, and the expression for $\alpha$ may be found in \cite{Martelli:2013aqa}. For more details we refer to 
\cite{Martelli:2013aqa}. The associated Reeb vector field is 
\bea
K & = & b_1 \partial_{\varphi_1} +  b_2 \partial_{\varphi_2} ~,
\eea
where
\bea
b_1 \ = \  \left|\frac{\sqrt{\alpha}+p_4^2}{p_3^2-p_4^2} \mathcal{P}'(p_3)\right| ~,\quad \qquad b_2 \ = \ \left|\frac{\sqrt{\alpha}+p_3^2}{p_3^2-p_4^2}\mathcal{P}'(p_4) \right|~. 
\label{rebel}
\eea
Remarkably,  using the relations given in \cite{Martelli:2013aqa}, we find that 
\bea\label{bob}
\sqrt{\frac{b_1}{b_2}}  \  = \   \beta^\mathrm{MP} (a,v)~,
\eea
in the three families of backgrounds considered in that reference. 

Equation (\ref{bob}) agrees perfectly with our field theory result. In particular, 
this result also gives an elegant explanation for why
certain non-trivial background geometries in \cite{Martelli:2012sz}, \cite{Martelli:2013aqa} 
had free energy equal to the round sphere result: it is simply because for those
backgrounds $b_1=b_2$, and hence $\beta=1$.
Moreover, recall that the general backgrounds in \cite{Martelli:2013aqa} led to an expression for $\beta^\mathrm{MP} (a,v)$ that generically is complex, and 
a corresponding complex large $N$ free energy. This indicates that our field theory analysis may be extended to the case where the background 
fields are complex. Although in these cases
the almost contact one-form bilinear $e^3$ is \emph{not} dual to a Killing vector field, if there exists another solution $\tilde \epsilon$ to the rigid Killing spinor equation with opposite R-charge (satisfying an analogue of equation (\ref{KSEconj})), 
then the bilinear $\tilde K = (\tilde \epsilon^{c})^\dagger \gamma^\mu \epsilon=\tilde{b}_1\partial_{\varphi_1}+
\tilde{b}_2\partial_{\varphi_2}$ yields two real Killing vectors \cite{Closset:2012ru}. It is natural to expect that on manifolds with three-sphere topology
the partition function will take the form we have found, where $\beta^2 = \frac{\tilde b_1}{\tilde b_2}$ is now a complex parameter.

\subsection{Vortex-antivortex factorization}\label{vortex}

In this section we would like to make contact with the results of \cite{Pasquetti:2011fj, Beem:2012mb}. The general metric considered in this paper has the form
\begin{equation}
\label{ours}
\dd s^2 \ = \ \Omega^2 (\dd\psi+a)^2+c^2 \dd z \dd\bar z~.
\end{equation}
We now specialize to the compact $S^3$ obtained by gluing together the two cigars\footnote{Note that $\beta$ in \cite{Beem:2012mb} is called $\alpha$ here.}
\begin{eqnarray}
\label{sara}
\dd s^2_{I} &=& \dd r^2+f(r)^2 \left(\dd\varphi_1+\epsilon \alpha \dd\varphi_2 \right)^2+\alpha^2 \dd\varphi_2^2~,\\
\dd s^2_{II} &=& \dd r^2+\tilde{f}(r)^2 \left(\dd\varphi_2+\tilde \epsilon \tilde \alpha \dd\varphi_1 \right)^2+\tilde \alpha^2 \dd\varphi_2^1 \nonumber~.
\end{eqnarray}
The cigars are glued at $r=\infty$, where we assume $f(r) \rightarrow \rho,~\tilde f(r) \rightarrow \tilde \rho$ as $r \rightarrow \infty$. Continuity of the metric across the gluing implies
\begin{eqnarray}
\tilde \alpha^2 \ = \ \frac{\rho^2}{1+\epsilon^2 \rho^2}~,~~~~~\tilde \epsilon^2 \ =\  \frac{\epsilon^2 \rho^2}{\alpha^2(1+\epsilon^2 \rho^2)}~,~~~~~\tilde \rho^2 \  =\  \alpha^2 (1+\epsilon^2 \rho^2)~.
\end{eqnarray}
Asymptotically the two metrics approach $\mathbb{R} \times T^2$, where the torus has complex structure $\tau= \epsilon \alpha+\ii \frac{\alpha}{\rho}$. One can explicitly check that the gluing takes $\tau \rightarrow \tilde \tau = \frac{1}{\tau}$.

In order to get from (\ref{ours}) to (\ref{sara}), let us perform the following change of coordinates
\begin{eqnarray}
z &=& g(r) \ex^{\ii (\beta^{-1} \varphi_1+ \beta \varphi_2 )}~,\nn\\
\bar{z} &=& g(r) \ex^{-\ii (\beta^{-1} \varphi_1+ \beta \varphi_2 )}~,\nn\\
\psi &=& \frac{1}{2\beta}\varphi_1 - \frac{\beta}{2}\varphi_2~.
\end{eqnarray}
In order to match (\ref{ours})  to $\dd s^2_I$ in (\ref{sara}) we must set
\begin{eqnarray}
a &=& \frac{(\beta^4- \epsilon^2\alpha^2)f(r)^2-\alpha^2}{2\left[\alpha^2+(\beta^2-\epsilon\alpha)^2 f(r)^2\right]}  (\beta^{-1} \dd\varphi_1+ \beta \dd\varphi_2 )~,\nn\\
c &=& \frac{\beta \alpha f(r)}{g(r)\sqrt{\alpha^2+(\beta^2- \epsilon\alpha)f(r)^2}}~,\nn\\
\Omega&=& \frac{\sqrt{\alpha^2+(\beta^2-\epsilon\alpha)f(r)^2}}{\beta}~,
\end{eqnarray}
where $g(r)$ is such that it satisfies the simple differential equation
\begin{equation}
c(r)^2 g'(r)^2 \ = \ 1~.
\end{equation}
Matching (\ref{ours})  to $\dd s^2_{II}$ we obtain similar expressions, related to the expressions above by the gluing conditions, provided one also takes $\beta \rightarrow \frac{1}{\beta}$. 
In order to make contact with the two-dimensional analysis of \cite{Beem:2012mb}, we must turn off the background field $V$. In the {\it real} case considered in this paper, this can be achieved by setting
\begin{equation}
\beta^2 \  =\  \epsilon \alpha~.
\end{equation}
In general we expect $\beta^2=\tau$, which is complex and becomes real only in the limit $\rho \rightarrow \infty$. Hence, in order to apply the results of this paper, we focus on this limit. In this case we obtain
\begin{equation}
\label{cigarsol}
a \ =\  -\frac{1}{2}(\beta^{-1} \dd\varphi_1+ \beta \dd\varphi_2 )~,~~~~~c \ = \ \beta \frac{f(r)}{g(r)}~,~~~~~\Omega \ =\  \frac{\beta}{\epsilon}~.
\end{equation}

On the other hand, we can also consider the particular case where the metric (\ref{ours}) reduces to the metric of the squashed $S^3_{b}$ sphere:
\begin{equation}
\label{squashed}
\dd s^2_{S^3_b} \ = \ f_b(\theta)^2 \dd\theta^2 +\frac{1}{b^2} \sin^2\theta \dd\varphi_1^2 + b^2 \cos^2 \theta \dd\varphi_2^2~,
\end{equation}
with $f_b(0)^2=1/b^2$ and $f_b(\frac{\pi}{2})^2=b^2$ fixed by regularity. Furthermore, we require $V_1=V_2=0$ or equivalently $\Omega$ constant. This is achieved by setting $\beta = b$ and 
\begin{equation}
\label{squashedsol}
a \ =\  -\frac{1}{2} \cos(2\theta)(\beta^{-1} \dd\varphi_1+ \beta \dd\varphi_2 )~,~~~~~ c \ = \  \frac{\cos \theta \sin \theta}{g(\theta)}~,~~~~~\Omega\ =\ 1~,
\end{equation}
where $g(\theta)$ satisfies $c(\theta)^2 g'(\theta)^2=f_b(\theta)^2$.

We can now choose a map from $\theta$ in (\ref{squashed}) to $r$ in (\ref{sara}), such that $\theta~ \in~ [0,\frac{\pi}{4}]$ is mapped to $r~ \in~ [0,\infty]$ in $\dd s^2_I$, and 
$\theta~ \in~ [\frac{\pi}{4},\frac{\pi}{2}]$ is mapped to $r~ \in~ [\infty,0]$ in $\dd s^2_{II}$. If we now start from (\ref{ours}) and consider a one-parameter family of backgrounds, with fixed $\beta$, which  interpolates between (\ref{squashedsol}) and (\ref{cigarsol}), then the results of our paper imply that the partition function depends only on $\beta$! In other words, we have shown that we can deform $S_\beta^3$ into the union of two copies of $\R^2 \times_{\beta}S^1$, glued by an $S$-transformation, while keeping the Reeb vector field and hence the partition function invariant. This proves factorization.


\section{Conclusions}\label{SecConclusions}

In this paper we have computed the localized partition function 
for a general class of $\mathcal{N}=2$ supersymmetric Chern-Simons-matter 
theories on a three-manifold with the topology of a sphere. More precisely, we have studied 
the case in which the background geometry has real-valued 
fields $A$, $V$ and $h$, which implies the existence of 
two supercharges $\epsilon$, $\epsilon^c$, with opposite R-charge, 
and a Killing Reeb vector field $K=\partial_\psi$. 
Writing
this vector field in terms of the standard torus action $U(1)\times U(1)$ on $S^3$, via $K=
b_1\partial_{\varphi_1}+b_2\partial_{\varphi_2}$, then
we have shown that the partition function takes the form 
\bea\label{Zagain}
Z \, = \, \int \diff\sigma_0 \, \ex^{\frac{\ii\pi k}{b_1b_2}\mathrm{Tr}\, \sigma_0^2}\prod_{\alpha\in \Delta_+} 4
\sinh \frac{\pi\sigma_0\alpha}{b_1}\sinh\frac{\pi\sigma_0\alpha}{b_2}\prod_{\rho}s_\beta\left[\frac{\ii Q}{2}(1-\rc)-\frac{\rho(\sigma_0)}{\sqrt{b_1b_2}}\right]~,
\eea
and depends on the background geometry only through the 
single parameter $\beta=\sqrt{b_1/b_2}$.  This includes all previously 
known examples studied in the literature.
The large $N$ limit of this 
partition function agrees with the gravitational free energy 
for the two-parameter family of solutions constructed 
in \cite{Martelli:2013aqa}, which in turn generalizes 
earlier results in \cite{Martelli:2011fu, Martelli:2011fw, Martelli:2012sz}. 
Even though our field theory result was derived for real backgrounds, and hence real $\beta$,
as described at the end of section \ref{LargeNLimit} we 
conjecture (\ref{Zagain}) analytically continues to complex 
values of $\beta$. In particular, this result could then be used to prove 
vortex-antivortex factorization in the case with general complex $\tau$, as discussed 
in section \ref{vortex}.

Following the methods of this paper, one could also compute
the partition function for more general three-manifolds. 
In particular, it would be interesting to consider 
the case where $\pi_1(M_3)$ is infinite, the simplest example 
being $S^1\times S^2$.  We expect the computation of the localizated partition function to be radically 
alterered in this case. 
In particular, in this setting it is more appropriate to study the supersymmetric 
index. Finally, it would be interesting, and straightforward, to extend our results to include the 
localization of BPS observables, for example Wilson loops and vortex loops \cite{Drukker:2012sr}.

\subsection*{Acknowledgments}
\noindent We would very much like to thank Stefano Cremonesi and Achilleas Passias for collaboration on 
part of this project. D.~M. and J.~F.~S. would like to thank Francesco Benini for discussion on
related material. 
 The work of L.~F.~A and P.~R. is supported by ERC STG grant 306260. L.~F.~A. is a Wolfson Royal Society Research Merit Award holder. D.~M. is supported by an ERC Starting Grant - Grant Agreement N. 304806 - Gauge-Gravity, and also acknowledges partial support from an EPSRC Advanced Fellowship EP/D07150X/3 and from the STFC grant ST/J002798/1. J.~F.~S. is supported by a Royal Society Research Fellowship and Oriel College, Oxford.

\appendix

\section{Conventions}\label{AppConventions}

In this appendix we present more details of our conventions. 
Greek indices $\mu,\nu=1,2,3$ denote spacetime indices while $a,b=1,2,3$ denote indices in an orthonormal frame. In the orthonormal frame (\ref{frame})
defined in the main text the gamma matrices are taken to be 
\bea
\gamma_1 &=&  -\sigma_1~, \qquad \gamma_2 \ =\  -\sigma_2~, \qquad\gamma_3 \ =\  \sigma_3~,
\eea
($\sigma_a$, $a=1,2,3$, denote the Pauli matrices) which follow from the conventions in \cite{Closset:2012ru} after cyclically permuting their orthonormal frame as $\{ 1,2,3\}_{\mathrm{CDFK}} \rightarrow \{3,1,2 \}$. These gamma matrices obey
\bea
\gamma_a \gamma_b & =&  \delta_{ab} + \ii \epsilon_{abc}\gamma_c~.
\eea
Spinors on the three-manifold $M_3$ are taken to be commuting/Grassmann even.  The charge conjugate of a spinor $\epsilon$ is defined  as\footnote{In particular in the real case that we 
study this means that the spinor and tilded spinor in \cite{Closset:2012ru} are related as $\tilde{\zeta} = -\ii \zeta^c$.} 
\bea
\epsilon^c &\equiv & \sigma_2\epsilon^*~,
\eea
The Fierz identity for commuting spinors is
\bea
\chi \psi^\dagger  &= & \frac{1}{2} \left( (\psi^\dagger \chi ) {\bf 1} +(\psi^\dagger \gamma_\mu \chi) \gamma^\mu \right)~, 
\eea
where  $\dagger$ denotes the usual Hermitian conjugate.

\section{Further background geometry}\label{AppMoreFormulae}

We begin by recording the inverse frame to (\ref{frame}). Writing 
\bea
z & =& x_1+\ii x_2~, \qquad a \ = \ a_{x_1}\diff x_1 + a_{x_2}\diff x_2~,
\eea
 we have
\bea\label{inverseframe}
\partial_1 &=& \frac{1}{c}\partial_{x_1} - \frac{a_{x_1}}{c}\partial_\psi~,\nn\\
\partial_2 &=& \frac{1}{c}\partial_{x_2} - \frac{a_{x_2}}{c}\partial_\psi~,\nn\\
\partial_3 &=& \frac{1}{\Omega}\partial_\psi~.
\eea
The spin connection associated with our orthonormal frame is defined as usual through 
\bea
\omega_{\mu a}{}^b & = & - e^b_\nu \nabla_\mu e_a^\nu~,
\eea
where the covariant derivative acts on $e_a^\nu$ as a vector, which leads to
\bea
\omega^{12} & =& \partial_2\log c\, e^1 - \partial_1\log c\, e^2 - \frac{\Omega}{2c}(\partial_1a_{x_2} - 
\partial_2a_{x_1}) e^3~,\nn\\
\omega^{13} &=& -\frac{\Omega}{2c}(\partial_1 a_{x_2}- \partial_2 a_{x_1})e^2 - \partial_1\log\Omega \, e^3~,\nn\\
\omega^{23} &=& \frac{\Omega}{2c}(\partial_1a_{x_2}-\partial_2 a_{x_1})e^1 - \partial_2\log\Omega\, e^3~.
\eea

The covariant derivative of a spinor $\psi$ is
\bea
\nabla_\mu \psi &=& \left( \partial_\mu + \frac{\ii}{4} \omega_{\mu ab} \epsilon^{abc} \gamma_c \right) \psi~.
\eea

We may derive an expression for the background gauge field $A$ in terms of the other background fields and the spin connection. Starting from the nowhere vanishing one-form $P_\mu = \ii \epsilon^{c \dagger} \gamma_\mu \epsilon = s ( e^1 + \ii e^2 )_\mu$ and differentiating we find
\bea
\nabla_\mu P_\nu &=& 2 \ii A_\mu P_\nu - \epsilon_{\mu \nu \rho} P^\rho h - 2 \ii V_\mu P_\nu - \ii V_\nu P_\mu + \ii g_{\mu \nu} V^\rho P_\rho ~.
\eea
From this we form 
\bea
\bar{P}^\nu \nabla_\mu P_\nu &=&  2 \ii A_\mu |P|^2 - \epsilon_{\mu \nu \rho} \bar P^{\nu} P^\rho h - 2 \ii V_\mu |P|^2 + \ii \left( V \cdot P \bar{P}_\mu - V \cdot \bar{P} P_\mu \right) \label{Peq_1} \, ,
\eea
where $\bar{P}_\mu = \bar{s} ( e^1 - \ii e^2 )_\mu$. The real and imaginary part of \eqref{Peq_1} give
\bea
\bar{P}^\nu \nabla_\mu P_\nu + P^\nu \nabla_\mu \bar P_\nu \ = \ \nabla_\mu |P|^2 &=& 2 \ii \left( V \cdot P \bar P_\mu - V \cdot \bar P P_\mu \right)~,\label{Aeq_1}\\
\bar{P}^\nu \nabla_\mu P_\nu - P^\nu \nabla_\mu \bar P_\nu &=& 4 \ii A_\mu |P|^2 - 2 \epsilon_{\mu \nu \rho} \bar P^{\nu} P^\rho h - 4 \ii V_\mu |P|^2 ~.\label{Aeq_2}
\eea
As $|P|^2= 2 \Omega^2$, equation \eqref{Aeq_1} in the frame yields
\bea
\partial_1 \Omega &=& -V_2 \Omega ~, \qquad \partial_2 \Omega=V_1 \Omega ~,
\eea 
whilst \eqref{Aeq_2} gives
\bea
A_\mu &=& \frac{1}{2} h e^3_\mu + V_\mu + j_\mu~,
\eea
with 
\bea
j_\mu &=& \frac{\ii}{4 \Omega^2} \left(s \partial_\mu \bar s - \bar s \partial_\mu s \right) + \frac{1}{2} \omega_\mu{}^1{}_2 ~.
\eea

\end{document}